# In Quest of a Ferromagnetic Insulator - Structure Controlled Magnetism in Mg-Ti-O Thin Films

Johannes Frantti*[,1], Yukari Fujioka[1], Christopher Rouleau[2], Alexandra Steffen[3], Alexander Puretzky[2], Nickolay Lavrik[2], Ilia N. Ivanov[2] and Harry M. Meyer[2]

[1]Finnish Research and Engineering, Helsinki 00180, Finland

[2]Center for Nanophase Materials Sciences, Oak Ridge National Laboratory, Oak Ridge, Tennessee 37831, USA

[3]Neutron Scattering Division, Oak Ridge National Laboratory, Oak Ridge, Tennessee 37831, USA

*Corresponding author: e-mail johannes.frantti@fre.fi

## Abstract

Ferromagnetic insulator thin films can convey information by spin waves, avoiding charge displacement and Eddy current losses. The sparsity of high-temperature insulating ferromagnetic materials hinders the development of spin wave based devices. Stoichiometric magnesium titanate, $MgTiO_3$, has an electronic-energy-band structure in which all bands are either full or empty, being a paramagnetic insulator. The $MgTiO_3$ ilmenite consists of ordered octahedra and cation network in which one third of the octahedra are vacant, one third host magnesium and one third titanium. By giving up these characteristics, a rich variety of different magnetic structures can be formed. Our experiments and electronic-energy-band-structure computations show that the magnetic and electric properties of Mg-Ti-O films can drastically be changed and controlled by Mg- and Ti-cation arrangement and abundancy in the octahedra. Insulating titanium- and semiconducting magnesium-rich films were ferromagnetic up to elevated temperatures. The presence and origin of ferromagnetic insulating phase in the films is not apparent - the expectation, based on the well-established rules set by Goodenough and Kanamori, is paramagnetic or antiferromagnetic ordering. We show that ferro- and paramagnetic phases, possessing the same stoichiometry, can be obtained by merely rearranging the cations, thus allowing defect-free interfaces in multilayer structures.

## Introduction

Interest towards electrically insulating ferromagnetic material is considerable in devices utilizing spin-wave based information processing. In contrast to metallic ferromagnets, insulating ferromagnetic oxides do not suffer from Eddy currents. Eddy currents cause power loss, and are a key mechanism underlying spin-wave damping. There is a surplus of ferrimagnetic compounds, such as $Y_3Fe_5O_{12}$ (YIG). In contrast ferromagnetic insulators are rare since antiferromagnetic ordering prevailingly has an energy advantage over the ferromagnetic ordering [1]. The well-known ferromagnetic insulators include EuO and EuS, which have Curie temperatures 77 and 16K, respectively. Rare earth elements and Cr, Mn, Co, Ni, and Fe are common cation candidates when new magnetic materials are developed. The lack of 3*d* electrons in Mg makes it overlooked in magnetic compounds. Ti has two 3*d*-electrons, which typically participate on bond forming, and rarely are ferromagnetically coupled. As is shown below, this general view is not a requirement set by nature and Ti can host unpaired electrons from Mg, which become ferromagnetically coupled.

Stoichiometric magnesium titanate, $MgTiO_3$, has the rhombohedral ilmenite structure [2]. $MgTiO_3$ can be grown as high-quality thin films with high-quality factor and a large optical band gap of approximately 4eV [3]. As a well-characterized dielectric material, $MgTiO_3$ thin films find applications in microwave applications [4] and optical devices [5]. Stoichiometric $MgTiO_3$ is a paramagnet as all electrons are paired. Cations with unpaired electrons, such as Co and Ni, are often substituted for Mg to improve the dielectric properties [3,6]. The substitutions are typically of the order of 5 at.% and thus do not introduce long-range magnetic ordering. $Ti_2O_3$ stabilizes to the corundum structure and is a classical material exhibiting metal-insulator transition (MIT). $Ti^{3+}$ has one unpaired 3*d* electron, which results in paramagnetic contribution above MIT, and forms a covalent bond with another $Ti^{3+}$ cation below the transition. The distance between the two $Ti^{3+}$ cations is shortest through the shared octahedron face, and thus the bond is parallel to the hexagonal *c*-axis. Essential structural characteristic is the short distance between two Ti, 2.8Å. The cations are not intervened by oxygen. Due to the electron pairing no local magnetic moments are expected in $Ti_2O_3$, consistently with experimental observations [7]. In magnesium titanate unpaired electrons and further ferromagnetic order can be introduced by changing the Mg:Ti ratio. In a bulk $Mg_{1-x}Ti_{1+x}O_3$ the formation of titanium-rich magnesium titanate results in ferromagnetic ordering stable up to 260 K [8]. Electronic-energy-band structure computations indicated that excess Ti at the Mg-O cation layer results in a partially occupied band at the conduction band edge: the lowest part of the conduction band edge corresponds to the spin-up states [8]. The structure is still the ilmenite in which one third of the octahedra are vacant and excess Ti has replaced Mg.

Here we show that in thin films another route for manipulating magnetic and electric properties is to vary octahedra filling: in thin films the cation distribution and filling fraction can be significantly different than in the ilmenite or the corundum structures. In thin films magnesium and titanium cations are mixed, and also the octahedra filling fraction is no longer constrained to be 2/3, i.e., the cation/oxygen at.% ratio is no longer 2/3. This serves as a way to manipulate magnetic properties and control specific resistivity. Crucially, it is possible to grow compositionally equivalent but magnetically completely different - paramagnetic and ferromagnetic - films with evident device applications.

## Methods

Mg-Ti-O thin films were deposited by PLD technique at the Center for Nanomaterials Sciences on sapphire substrates with 0001 (C-plane), $11\bar{2}0$ (R-plane) and $10\bar{1}0$ (M-plane) planes. $Ti_2O_3$ and MgO targets were ablated repeatedly [9]. Shot ratios and the total number of shots were adjusted to control the composition and film thickness and the targets were rotated during deposition. Both Mg- and Ti-rich films were deposited. Substrates were in situ heated to 923 K. Laser beam wavelength was 248nm, pulse repetition rate 10Hz, and fluence between 3.1 and 3.2$Jcm^{-2}$.

High-resolution Raman measurements were performed using a Jobin-Yvon T64000 spectrometer consisting of a double monochromator coupled to a third monochromator stage with 1800 groves/mm grating (double subtractive mode). A liquid nitrogen cooled charge-coupled device detector was used to count photons. The high-intensity Raman spectra were measured in a custom micro-Raman setup using 532nm excitation and a 100×-microscope objective with NA (numeric aperture)=0.9 (beam spot on the samples ≈1μm). All measurements were carried out in backscattering geometry.

X-ray diffraction (XRD) data on thin films were collected by Panalytical X'Pert Pro MPD diffractometer at the CNMS X-ray diffraction laboratory. A CuKα source, Ni filter and X'Celerator detector were used in the measurements. Film thicknesses were estimated by JA Woollam M-2000U ellipsometer. When possible, the thicknesses were also estimated from the subsidiary maxima in the XRD patterns. Composition and valence states as a function of depth were determined by X-ray photoelectron spectroscopy (XPS) measurements (XPS, K-Alpha XPS system, Thermo Fisher Scientific) equipped with a monochromated Al-Ka source (hν=1486.6eV) and argon ion sputtering. Table 1 gives a sample summary and the film compositions determined by XPS measurements. It also tabulates parameters *n*, *v* and $z_v$ defined below. The thicknesses of the $Mg_{0.73}Ti_{1.38}O_3$, $Mg_{1.01}Ti_{1.27}O_3$, $Mg_{1.24}Ti_{1.17}O_3$, $Mg_{1.49}Ti_{1.03}O_3$ and $Mg_{1.64}Ti_{0.98}O_3$ films used in magnetic measurements were 63, 62, 62, 51 and 47nm, respectively, and the lateral dimensions were 5mm×5mm.

Table 1. Samples. The composition is given in terms of the formula $Mg_{1+nx}Ti^{v}_{1-x}O_3$. The fraction of the filled octahedra is labelled as $z_v$. Mg valence was +2 and XPS revealed that Ti has mixed valence states, and *v* is the average Ti valence computed from the experimentally determined composition and charge neutrality condition. *M* is the saturation magnetization value.

| Sample | Composition | Substrate | *x* | *n*(*x*) | *v*(*x*) | $z_v(x)$ | *M* ($emucm^{-3}$) |
|---|---|---|---|---|---|---|---|
| M1 | $Mg_{1.49}Ti_{1.03}O_3$ | $Al_2O_3$ (0001) | -0.03 | -15.29 | 2.92 | 0.83 | 0 |
| M4 | $Mg_{1.24}Ti_{1.17}O_3$ | $Al_2O_3$ (0001) | -0.17 | -1.35 | 3.00 | 0.80 | 43 |
| M5 | $Mg_{1.24}Ti_{1.17}O_3$ | $Al_2O_3$ $(10\bar{1}0)$ | -0.17 | -1.35 | 3.00 | 0.80 | - |
| M7† | $Mg_{1.86}Ti_{0.90}O_3$ | $Al_2O_3$ (0001) | - | - | - | - | - |
| M8† | $Mg_{1.86}Ti_{0.90}O_3$ | $Al_2O_3$ $(10\bar{1}0)$ | - | - | - | - | - |
| M10 | $Mg_{1.01}Ti_{1.27}O_3$ | $Al_2O_3$ (0001) | -0.27 | -0.04 | 3.13 | 0.76 | 5 |
| M16,M22 | $Mg_{1.64}Ti_{0.98}O_3$ | $Al_2O_3$ (0001) | 0.02 | 33.2 | 2.78 | 0.87 | 9 |
| M19 | $Mg_{0.73}Ti_{1.38}O_3$ | $Al_2O_3$ (0001) | -0.38 | 0.69 | 3.27 | 0.70 | 12 |
| M20,M25 | $Mg_{0.73}Ti_{1.38}O_3$ | $Al_2O_3$ $(10\bar{1}0)$ | -0.38 | 0.69 | 3.27 | 0.70 | - |
| M29‡ | MgO | $Al_2O_3$ (0001) | | | | | |

† Sample contains brookite phase and suggests that the Mg-content exceeds the solubility limit.
‡ Rocksalt structure

Magnetization measurements were performed by a VSM device (Quantum Design PPMS Dynacool VSM). Magnetization data as a function of applied field were measured at constant temperatures. For electrical measurements interdigitated electrodes were patterned by photolithography and a mask on samples grown with the same deposition parameters as the samples used in magnetic measurements. Electrodes were aligned differently in order to study possible orientation dependent conductivity. Also electron beam deposition with a shallow mask was applied for depositing surface electrodes. In this case a 5 nm thick Ti-layer was first deposited after which 100 nm thick Au-layer was grown.

We used the ABINIT DFT code for spin-polarized computations [10].The projector augmented wave [11] approach, implemented to Abinit code [12,13] was applied for the computations. The energy cutoff and the energy cut-off for the fine Fast-Fourier-Transform (FFT) grid were 50 and 100 Hartree, respectively, and the *k*-point meshes were 10 × 10 × 10 for the density-of-state computations. Structures were optimized prior to the electronic-band-structure computations. Ferromagnetic, ferrimagnetic and antiferromagnetic structures were tested for every structure to avoid local minima in the optimization stage. The minimum energy structures are given.

## Results and discussion

### XRD results

Fig. 1 shows xrd patterns measured from Ti-rich and Mg-rich films. In all patterns, except for the MgO (which has the rocksalt structure) and $Mg_{1.86}Ti_{0.90}O_3$ (possessed two phases) films, only 003, 006, and 0012 reflections were present. Reflections were indexed in terms of the hexagonal axes. The *c*-axes lengths, given in Table 2, are close to the value $c$ = 13.8992 (7) Å reported in ref. 2. The intensities of the 00*l* reflections were strongly composition dependent, and 009 and 0015 reflections were absent. The ilmenite structure allows 00*l* reflections, when *l* is an integer multiple of 3, and the corundum structure allows 00*l* reflections, when *l* is an even multiple of 3. The ilmenite and corundum structures are not consistent with the observed features. As discussed below, XPS measurements indicated that the cation content significantly exceeds the stoichiometric values characteristic to the ilmenite and corundum structures. Based on the XPS, XRD and Raman data a model in which also the 3*a* and 3*b* sites of the $R\overline{3}$ space group symmetry can be occupied was applied to explain the experimental results. As the straightforward computation given below indicates, filling 3*a* and 3*b* sites explains the absence of 009 and 0015 reflections.

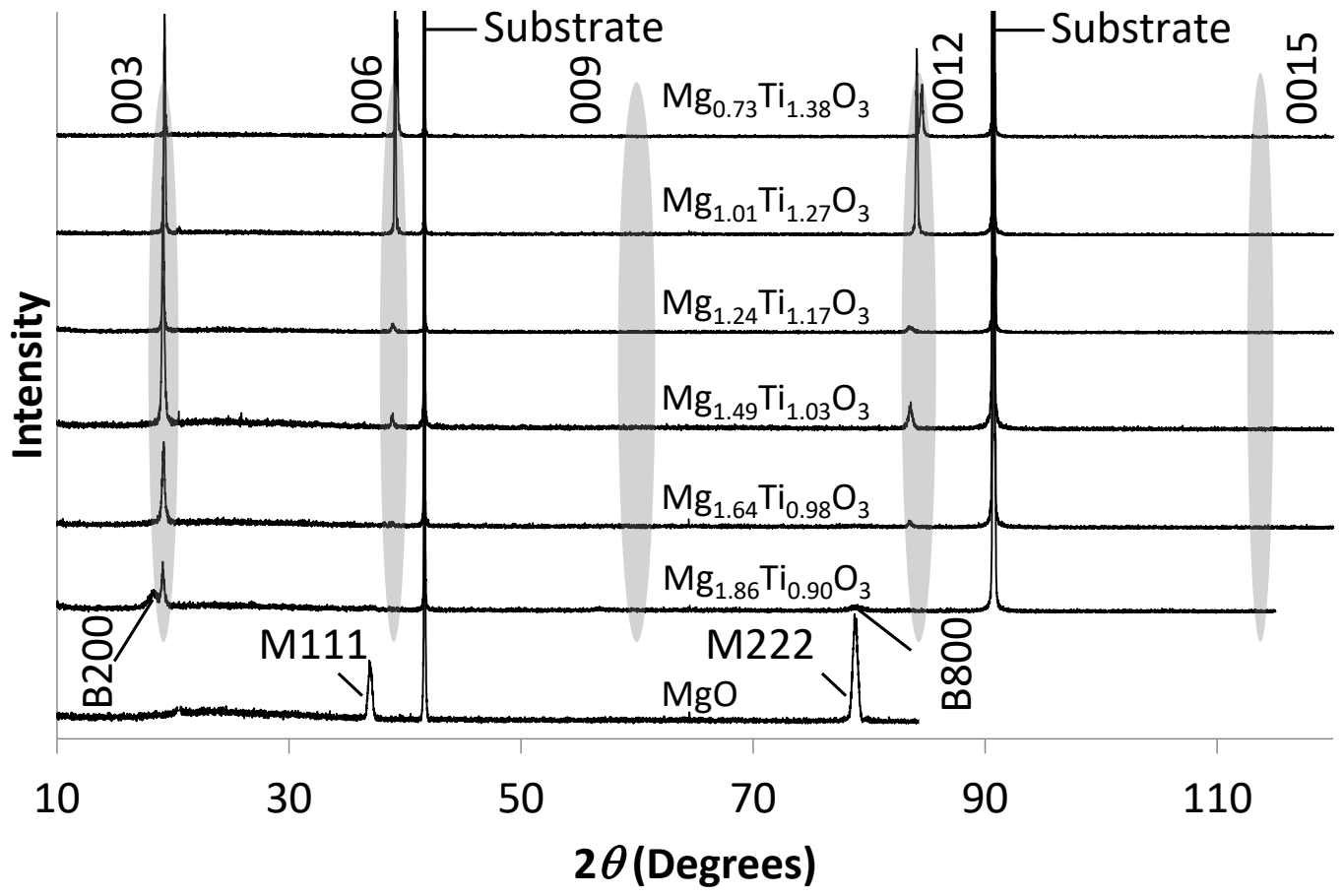

Fig. 1. X-ray diffraction patterns collected on thin films deposited on C-plane sapphire. Grey ovals highlight the 003, 006, 009, 0012 and 0015 reflection positions. The 009 and 0015 reflections have zero intensity. $Mg_{1.86}Ti_{0.90}O_3$ possesses both the brookite phase, labelled by B, and the ilmenite derived structure. The other patterns correspond to the single phase thin films with the *c*-axis perpendicular to the substrate plane. Pure MgO film has the rocksalt structure.

Fig. 2 shows the relative intensities of the 00*l* reflections as expected for the aforementioned model, together with the experimentally determined values. When the cations enter at 6*c*, 3*a* and 3*b* sites and oxygen at 18*f* site, the structure factor $S_{00l}$ can be written as

$$S_{00l} = 2f_A \cos 2\pi l z_1 + 2f_B \cos 2\pi l z_2 + f_C + f_D(-1)^l + 3f_O \cos 2\pi l z_3$$

where the cation site occupancy is between 0 and 1 and each cation site can hold an average scatterer, consisted of a fraction of $Ti^{3+}$ and $Mg^{2+}$ cations. In this case the structure factors possess no imaginary part and the intensity of the 00*l* reflection is directly proportional to the square of the structure factor. Atomic form factors$f_A$,$f_B$,$f_C$,$f_D$and$f_O$ refer to cation scatterers at sites 6*c*(00$z_1$), 6*c*(00$z_2$), 3*a*(000), 3*b*(00½) and oxygen at site 18*f*($x_3y_3z_3$), respectively. In the present case, the observed peak intensities are independent from the fractional coordinates $x_3$ and $y_3$. Atomic form factors for cations and oxygen were estimated from equation $f(\sin\theta/\lambda) = \sum_{i=1}^{4} a_i \exp(-b_i \sin^2\theta/\lambda^2) + c$ [13]. Parameters $a_i$, $b_i$ and $c$ were taken from ref. 14. When the same site shares both $Mg^{2+}$ and $Ti^{3+}$ cations in atomic percentage fractions *x* and 1-*x*, respectively, the atomic form factor was taken as $xf_{\mathrm{Mg}} + (1-x)f_{\mathrm{Ti}}$. Polarization, Lorentz and geometrical factors were modelled by multiplying the square of the structure factor by a function $g(\theta) = (1+\cos^2 2\theta)/(\sin^2\theta\cos\theta)$. Table 2 tabulates the fractional *z*-coordinates and site occupancies for five single phase thin films. The cation and oxygen ratios were determined by XPS measurements.

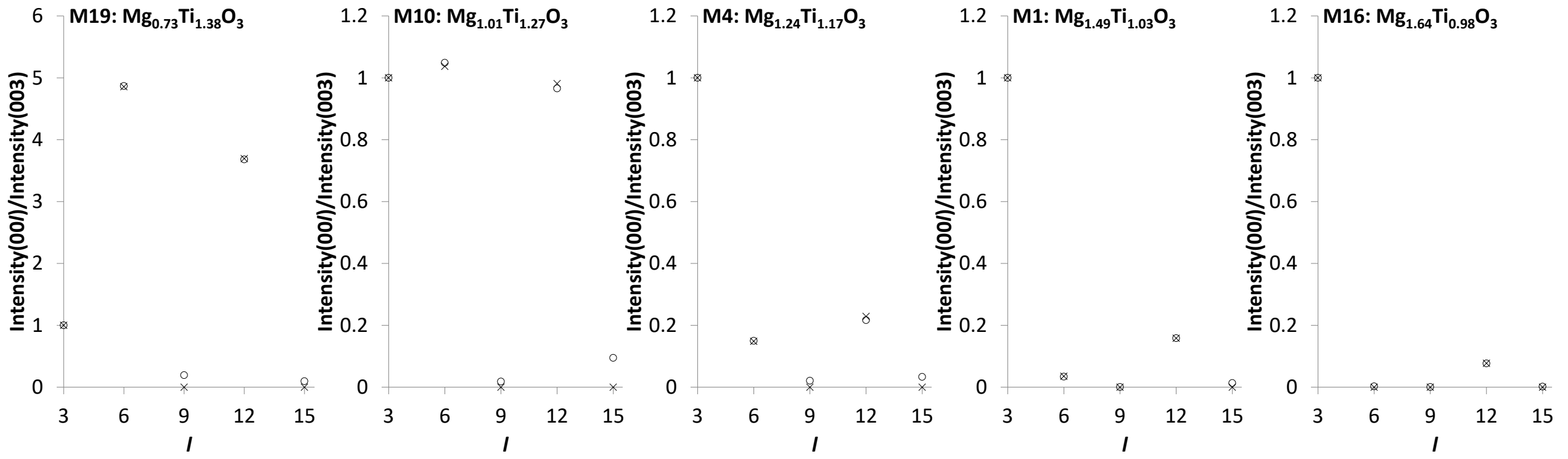

Fig. 2. Intensities of the 00*l* reflections relative to the 003 reflection. Experimental data are presented by crosses and fitted data by circles. Note the different intensity scale in the left-most panel.

Table 2 shows that Mg and Ti atoms prefer to segregate on different layers perpendicular to the *c*-axis, which is reminiscent to what is found in the ilmenite $MgTiO_3$. This is also seen from the cation distribution at 3*a* and 3*b* sites. An exception is the $Mg_{0.73}Ti_{1.38}O_3$ film, in which the cations are more evenly distributed over the sites. Despite the relatively low in-situ substrate annealing temperatures, the preference of 6*c*-site occupancy filling over the 3*a* and 3*b* sites is seen. The excess cations were placed to the 3*a* and 3*b* sites only after the 6*c* sites were filled. Fig. 2 compares the experimental and computed intensities and indicates that the given structural model is able to explain the reflection intensity behavior seen in Fig. 1.

Table 2. Cation and oxygen displacements and site occupancies adjusted so that the experimental intensities of the 003, 006, 009, 0012 and 0015 reflections are reproduced. SO refers to the site occupancy, and *z* to the fractional *z*-coordinate.

| Site | *x* | *y* | *z* | M16 $Mg_{1.64}Ti_{0.98}O_3$ $c$ = 13.877Å | | M1 $Mg_{1.49}Ti_{1.03}O_3$ $c$ =13.880Å | | M4 $Mg_{1.24}Ti_{1.17}O_3$ $c$ =13.872Å | | M10 $Mg_{1.01}Ti_{1.27}O_3$ $c$ =13.801Å | | M19 $Mg_{0.73}Ti_{1.38}O_3$ $c$ =13.746Å | |
|---|---|---|---|---|---|---|---|---|---|---|---|---|---|
| | | | | *z* | SO | *z* | SO | *z* | SO | *z* | SO | *z* | SO |
| 6*c* Mg | 0 | 0 | $z_1$ | 0.369 | 0.864 | 0.363 | 1 | 0.374 | 0.984 | 0.373 | 0.938 | 0.375 | 0.346 |
| 6*c* Ti | 0 | 0 | $z_1$ | 0.369 | 0.136 | 0.363 | 0 | 0.374 | 0.016 | 0.373 | 0.061 | 0.375 | 0.654 |
| 6*c* Mg | 0 | 0 | $z_2$ | 0.126 | 0.136 | 0.125 | 0 | 0.124 | 0.016 | 0.131 | 0.061 | 0.124 | 0.346 |
| 6*c* Ti | 0 | 0 | $z_2$ | 0.126 | 0.844 | 0.125 | 1 | 0.124 | 0.984 | 0.131 | 0.938 | 0.124 | 0.654 |
| 3*a* Mg | 0 | 0 | 0 | 0 | 0.312 | 0 | 0 | 0 | 0.120 | 0 | 0.016 | 0 | 0.038 |
| 3*a* Ti | 0 | 0 | 0 | 0 | 0 | 0 | 0.06 | 0 | 0.211 | 0 | 0.430 | 0 | 0.072 |
| 3*b* Mg | 0 | 0 | ½ | ½ | 0.968 | ½ | 0.98 | ½ | 0.360 | ½ | 0.004 | ½ | 0.038 |
| 3*b* Ti | 0 | 0 | ½ | ½ | 0 | ½ | 0 | ½ | 0.129 | ½ | 0.110 | ½ | 0.072 |
| 18*f* O | $x_3$ | $y_3$ | $z_3$ | 0.255 | 1 | 0.268 | 1 | 0.257 | 1 | 0.288 | 1 | 0.257 | 1 |

## Raman results

The Brillouin zone center modes of the prototype ilmenite structure (space group $R\bar{3}$) and corundum (space group $R\bar{3}c$) structure transform as the representation $5A_g$(R)+$5E_g$(R)+$4A_u$(IR)+$4E_u$(IR)+$A_u$(A)+$E_u$(A) and $2A_{1g}$(R)+$3A_{2g}$+$5E_g$(R)+$2A_{1u}$+$2A_{2u}$(IR)+$4E_u$(IR)+$A_{2u}$(A)+$E_u$(A), respectively. R, IR and A in braces indicate Raman and infrared active and acoustic modes, respectively. If the symmetry is kept the same the number of Raman active modes is not changed with changing octahedra filling fraction, though the intensities and mode frequencies change [15]. Filling 3*a* and 3*b* sites increases the number of silent and IR active modes. From the Raman scattering viewpoint, the most straightforward difference accompanied by a symmetry change $R\bar{3} \longrightarrow R\bar{3}c$ is the decrease of totally symmetry modes from 5 to 2, whereas the number of $E_g$ symmetry modes is preserved.

Study on Ni-Co-Ti-O thin films showed that notably the intensities undergo rather strong variation: with increasing filling fraction the peak at around 600 cm$^{-1}$ strengthens and the low-frequency peaks get weaker [15]. When the filling fraction of octahedra is close to 1, a symmetry change from rhombohedral symmetry to hexagonal $P6_3/mmc$ symmetry with one Raman active mode at around 600 cm$^{-1}$ was observed. The mode was assigned to the $E_g$ symmetry mode, which is the only Raman active mode allowed in the hexagonal phase.

Fig. 3(a) shows room Raman spectra and 3(b) the peak positions as a function of Ti/Mg ratio. As a reference, peak positions corresponding to the positions found in the ilmenite $MgTiO_3$ [16] and corundum $Ti_2O_3$ are given. The symmetry assignment follows the interpretation given in ref. 17 for data reported in ref. 18.

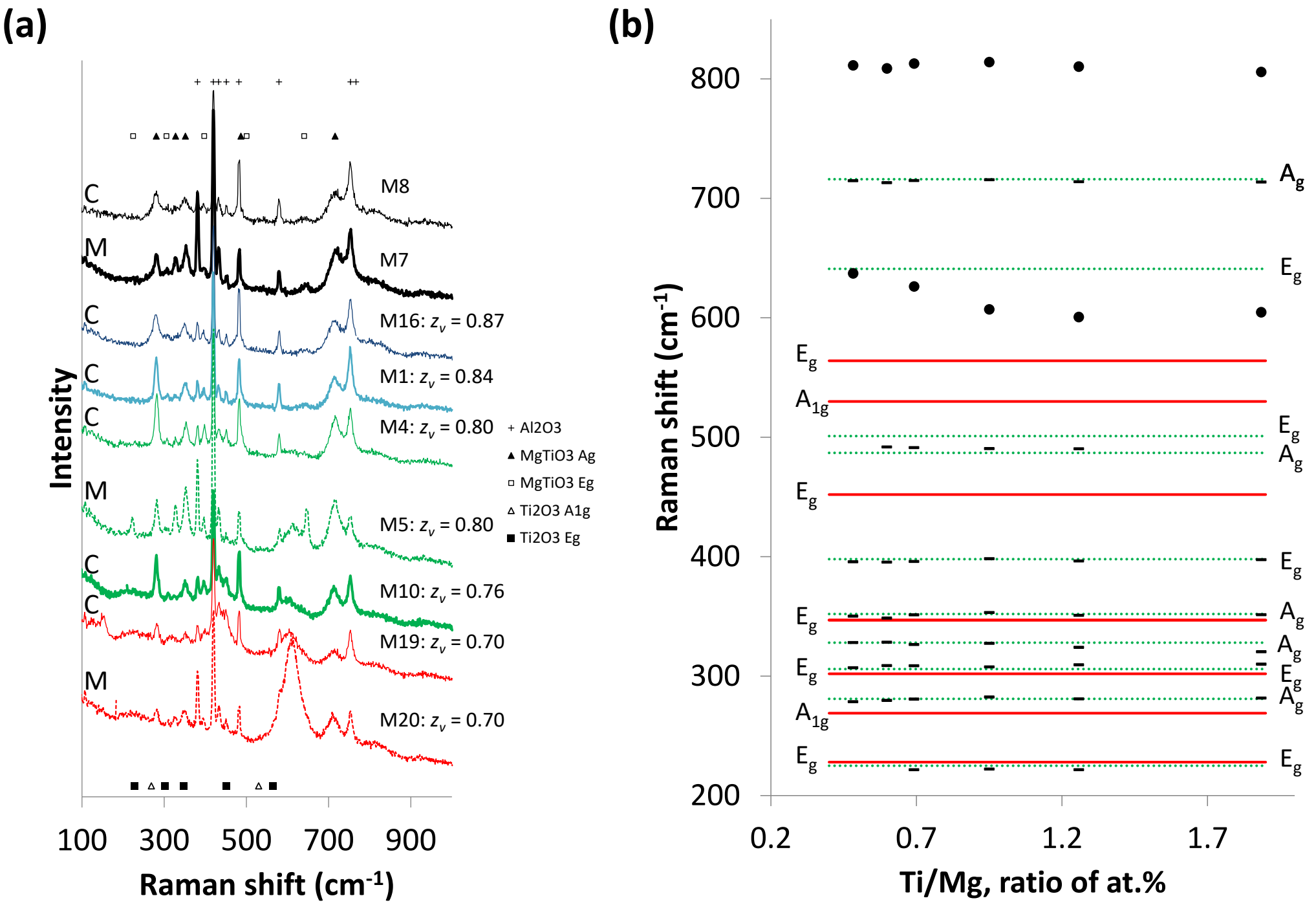

Fig. 3. (a) Room-temperature Raman spectra. The crosses indicate the peaks from the substrate. Note the change in the low-frequency range and the strengthening of the peak at around 600cm$^{-1}$. The filled triangles and open squares show the values reported for $MgTiO_3$ in ref. 16, and the open triangles and filled squares show the values reported for $Ti_2O_3$ [19]. Letters C and M refer to C- and M-plane of the substrate, respectively. (b) Raman shifts as a function of Ti/Mg ratio. The green dotted lines correspond to the values of $MgTiO_3$ [16] and the red continuous lines are the shifts reported for $Ti_2O_3$ [19]. Phonon symmetry labels on the left and right are for the $Ti_2O_3$ and $MgTiO_3$, respectively.

Comparison of the present data with the data given in ref. 16 indicates that the symmetry is better described by $R\bar{3}$ than $R\bar{3}c$, as Fig. 3(b) shows. The overall peak positions match well with the literature values, if one considers the decrease of the $E_g$ symmetry mode at 641cm$^{-1}$ with increasing Ti content. Due to the weak intensity and partial overlap with the substrate peak, it is not possible to determine whether there are one or two peaks originating from the sample in the region around 500cm$^{-1}$. The modest changes in the Raman peak numbers, considering the very large composition differences, are in line with the symmetry arguments given above. Fig. 3(a) shows that the intensity of the $E_g$ mode at around 600cm$^{-1}$ significantly strengthens with increasing Ti/Mg ratio, as is seen by comparing the spectra of the M1 and M20 samples. The highest frequency mode at around 800 cm$^{-1}$ does not correspond to the first-order Raman scattering and was rather weak and broad, see Fig. 3(a).

## XPS results

Fig. 4(a) shows the ratio of oxygen content to total cation content as a function of Ti/Mg ratio, extracted from XPS data collected at three depths. Also fits of data points to the expression (1) are given. Only valence parameter *v*, common to all data points corresponding to the same depth, was refined. Table 1 lists the valence values computed separately for each data point. Figs. 4(b)-(d) show XPS spectra measured from $Mg_{1.24}Ti_{1.17}O_3$ film. The dependence of average valence on etching time [*i.e.*, data sets are displaced vertically in Fig. 4(a)] is caused by preferential removal of O leaving the near-surface region chemically reduced.

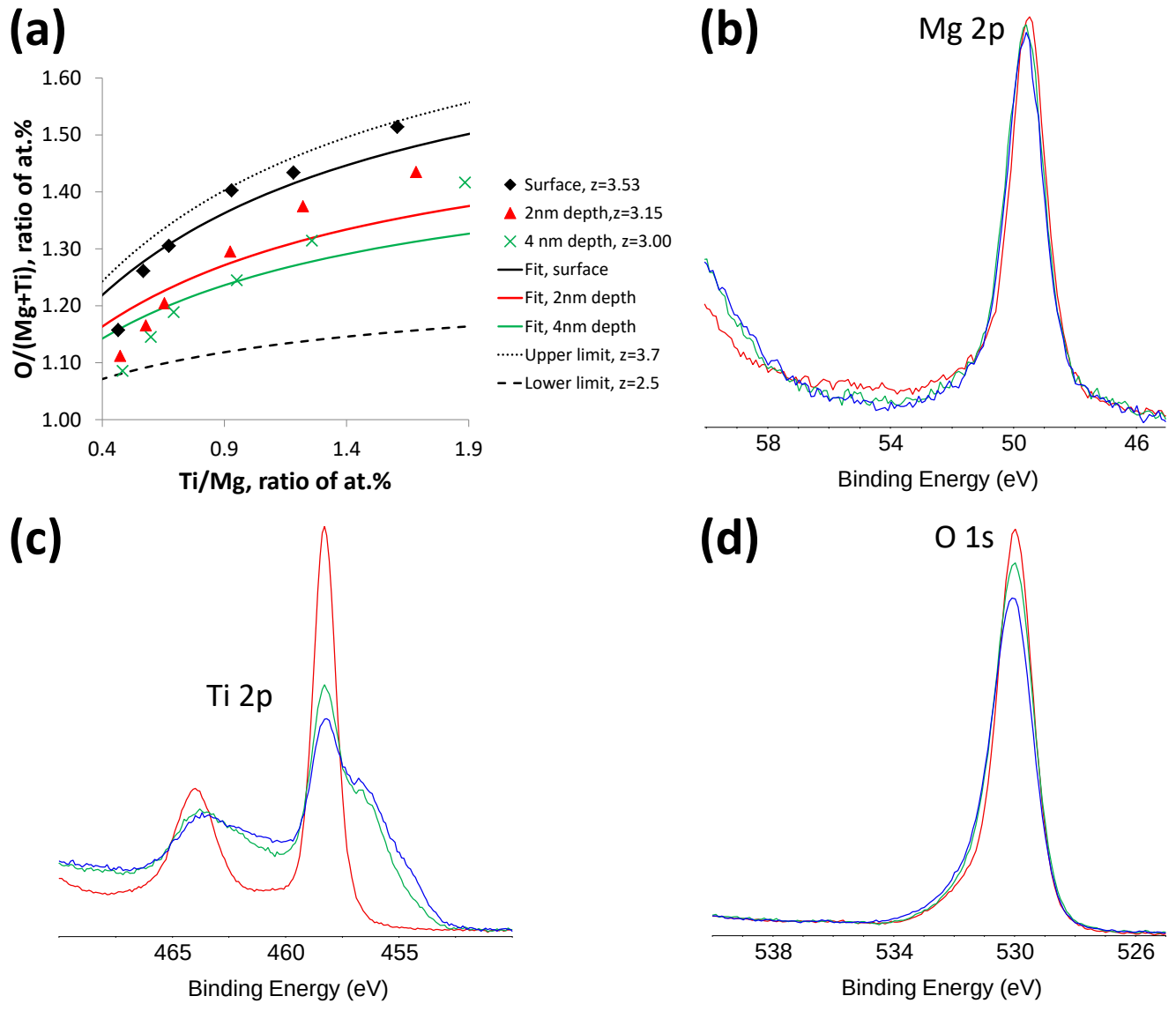

Fig. 4. (a) Ratio of oxygen content to total cation content as a function of Ti/Mg ratio as extracted from XPS data. Surface, 2nm, and 4nm refer to data collected on as-received surface and 2nm and 4nm etched surfaces, respectively. Panels (b)-(d) give Mg2*p*, Ti2*p* and O1*s* core level spectra measured from $Mg_{1.24}Ti_{1.17}O_3$ film for the as received (red line), 15 sec etched (green line), and 30 sec etched data (blue line), respectively.

## Structural model

Xrd measurements showed that single phase structure was formed in a large composition range, which with charge neutrality condition implies a change in the cation valence and/or structure. In stoichiometric $MgTiO_3$ Mg and Ti are in valence states +2 and +4, respectively, and Ti in $Ti_2O_3$ is in valence state +3. The corundum and ilmenite structures are consisted of oxygen octahedra of which 2/3 are filled. A structural model in which the fraction of filled octahedra changes as a function of Mg/Ti ratio gives a quantitative fit of the XPS data, Fig. 4. The film composition is expressed as

$$Mg^{2+}_{1+nx}Ti^{z}_{1-x}O_3 \qquad (1)$$

which is based on the following assumptions:

(1) Oxygen octahedra network is preserved, i.e., no major topological changes take place as a function of composition. According to xrd and Raman measurements this assumption is justified, except for the most Mg-rich samples $Mg_{1.86}Ti_{0.90}O_3$, which possessed the brookite phase as a secondary phase.
(2) Mg is in valence state 2+, which is justified by XPS measurements.
(3) Ti valence $v$ depends on composition, $v = v(x)$.
(4) A value for $nx$ is obtained from the charge neutrality condition: $nx=(4-v+vx)/2$.

The fraction of filled octahedra $z_v=[2+x(n-1)]/3$ is a convenient figure characterizing the crystal structure. If $v = 4$ then $n$ is 2 for all $x$ and Eq. (1) reduces to $Mg^{2+}_{1+2x}Ti^{v}_{1-x}O_3$. This, however, is not a sufficient approximation and, as listed in Table 1, $v$ is between 2.8 and 3.3 and correspondingly $n$ varies largely as a function of $x$. Thus, both the octahedra filling fraction and the Ti valence are variables preserving the charge neutrality. The $n$=2 case is found in (Ni,Co)-Ti-O oxides [15], where the physical interpretation of $n$ is straightforward, since $v(x)$=4 in the whole composition range and only $z_v$ changes. Thus, if one starts from a stoichiometric $(Ni,Co)^{2+}Ti^{4+}O_3$ ilmenite, all Ti-deficient compositions correspond to a substitution $Ni^{2+}/Co^{2+}$ for one $Ti^{4+}$ and insertion of another $Ni^{2+}/Co^{2+}$ in a vacant octahedron. Correspondingly, Ti-rich composition is achieved by substituting $Ti^{4+}$ for $Ni^{2+}/Co^{2+}$ and additionally removing one $Ni^{2+}/Co^{2+}$. In summary, changing Ti content by an amount $\Delta x$ is accompanied by a change $2\Delta x$ in 2+ cation content, hence $n$=2. In $Mg_{1+nx}Ti_{1-x}O_3$ similar approach is not valid, since Ti is in a 4+ state in the stoichiometric $MgTiO_3$ ilmenite, and the valence state of Ti changes as a function of $x$. This is seen from Fig. 4 which indicates that the constant valence fits deviate from the experimental data points at small and large Ti/Mg ratio values. $Mg_{1+nx}Ti_{1-x}O_3$ is better seen as a network of octahedra, where one inserts in average sense, (1-$nx$)/3 $Mg^{2+}$ cations and (1-$x$)/3 $Ti^{v}$ cations per octahedron.

Charge neutrality would also be preserved if all octahedra would be filled by $Mg^{2+}$, in analogy with the $(Ni^{2+}_{0.39}Co^{2+}_{0.61})_3O_3$ compound [15]. This limit may not be achievable; $Mg_{1.86}Ti_{0.90}O_3$ films (Table 1 and Fig. 1) have the brookite secondary phase. However, the structure can host rather large amounts of Mg, $Mg_{1.64}Ti_{0.98}O_3$ being so far the most Mg-rich single phase film.

As Fig. 4(a) shows, the oxygen content decreases with decreasing Ti/Mg ratio in all data sets, including the data collected on as-received surface, consistently with formula (1). In Mg-rich films the loss of high-valence (3<$v$<4) Ti-cations is compensated for by placing more $Mg^{2+}$ cations which (a) replace Ti and (b) enter to the vacant octahedra. The average Ti valence decreases with increasing etching time. When Ti/Mg is close to one, the as received spectrum has a binding energy position expected for Ti 4+, seen at 459 eV. The etched spectra show a mix of valences for the Ti (*i.e.* 2+, 3+, 4+), a feature which we discuss below in the context of electronic-energy-band computations. Fig. 4(a) gives the fits for each (surface, 2nm and 4 nm etched) data and upper and lower limits for the average valences. The upper and lower limit values correspond to curves which enclose all data points. A clear decrease in the relative oxygen fraction with decreasing Ti/Mg ratio is seen in Fig. 4(a) for constant etching times, consistently with the model outlines above. The fits shown in Fig. 4(a) deviate from the experimental data points to some extent: in Mg-rich areas experimental curves bend down faster than the model predicts, and vice versa in Ti-rich area. Fits illustrate the trends describing oxygen content as a function of Ti/Mg ratio, and a constant valence value was assumed in each fit. The deviation indicates that also valence value changes as a function of Ti/Mg ratio, as given in Table 1.

## Magnetic and electric measurements

Figure 5 shows hysteresis loops measured at 5 and 350K temperatures.

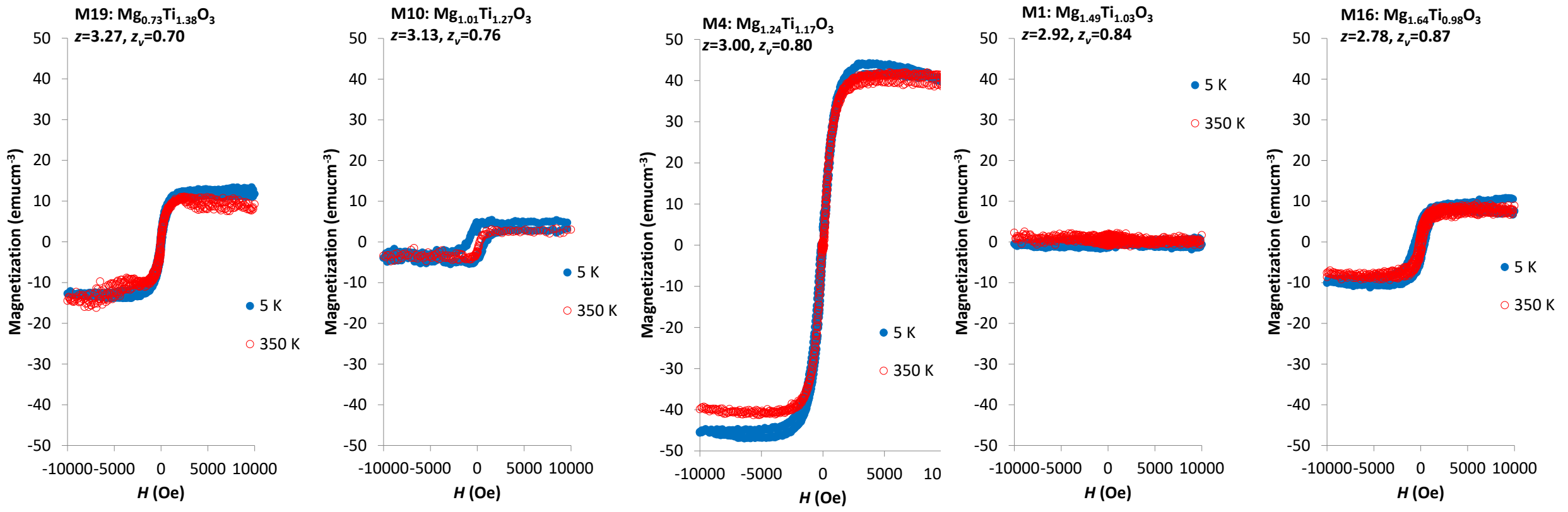

Fig. 5. In-plane magnetic hysteresis loops measured at two temperatures. A linear diamagnetic part due to the substrate was subtracted. Average valence $z$ and the octahedra filling fraction $z_v$ are given.

Sample $Mg_{1.24}Ti_{1.17}O_3$ has the largest magnetic moment, $43emucm^{-3}$. Table 1 gives experimentally determined magnetization values. Magnetization peaks at Ti/Mg ≈1.05 and has significantly smaller values for larger and smaller Ti/Mg ratio. In the stoichiometric corundum $Ti_2O_3$ the possible magnetization would be due to an unpaired electron of the Ti 3*d*-orbital derived state and, according the rules given in ref. 16, the ordering should be antiferromagnetic. If the electrons form a covalent bond along the *c*-axis, as is the case of $Ti_2O_3$, the material is paramagnetic. The essential feature of the present materials is that the 3*a* and 3*b* sites are occupied, Table 2. Paramagnetic behavior, found in $Mg_{1.49}Ti_{1.03}O_3$ film, is consistent with Table 2: the excess Mg-cations occupy the 3*b* site. The octahedra sites vacant in the Ti-O-layers of the ilmenite $MgTiO_3$ are filled by Mg, whereas the Mg-O layers remain approximately intact. Thus, the Ti-Mg-O layers are separated by diamagnetic Mg-O layers. The paramagnetic behavior is confirmed below by electronic-energy-band-structure computations. The situation is different in the other films, in which each layer perpendicular to the *c*-axis contains both Mg and Ti cations, Table 2. In bulk $Mg_{0.9}Ti_{1.1}O_3$ and $Mg_{0.88}Ti_{1.12}O_3$ samples excess Ti at the Mg-O cation layer results in a partially occupied band at the conduction band [8]. Similar explanation is valid in the present case, as the electronic-energy-band-structure computations, summarized below, indicate.

Though the magnetization values of the $Mg_{1.64}Ti_{0.98}O_3$ and $Mg_{0.73}Ti_{1.38}O_3$ samples are comparable, their specific resistance values are different, Table 3. Among the present samples, $Mg_{1.64}Ti_{0.98}O_3$ and $Mg_{0.73}Ti_{1.38}O_3$ exhibit the minimum and maximum value of $z_v$, respectively. Also other samples had resistance values comparable to sample $Mg_{0.73}Ti_{1.38}O_3$. Band-structure computations indicated that the role of cation disorder on the octahedral site plays a crucial role on resistivity of $V_2O_3$ [21].

Table 3. Specific resistance $\rho$ measured at 1Hz frequency.

| Sample | $\rho$ [MΩcm] | $z_v$ |
|---|---|---|
| M22: $Mg_{1.64}Ti_{0.98}O_3$ | 0.15 | 0.87 |
| M25: $Mg_{0.73}Ti_{1.38}O_3$ | 427 | 0.70 |

## Spin-polarization

In magnesium titanate the valence and conduction band edges are derived from the O2*p* and Ti3*d* states. Mg is ionically bonded to neighboring oxygen and serves as an electron donator. The dominating structural degrees of freedom found in the films are the cation distribution on the basal planes perpendicular to the *c*-axis, and the filling fraction of octahedra vacant in the ideal ilmenite structure, Table 2. To gain insight to the effect of cation distribution on the ground state magnetic properties, spin polarized computations were conducted on computationally tractable structures, derived from the $ABO_3$ ilmenite and $ABO_3$ lithium niobate structures. Both structures have the same octahedra network, the difference being the way the *A* and *B* cations are distributed. The case in which a fraction of Mg was replaced by Ti in the ilmenite structure results in ferromagnetism [8].

Fig. 6 summarizes the 9 nine structural variants derived from the ilmenite and the six variants derived from the lithium niobate structure with predicted magnetizations for each structure. Magnetization values are given in Bohr magneton units and correspond to the difference between the number of spin-up and spin-down electrons. The prototype ilmenite structure I1 and the hypothetical lithium niobate structure LN1 are paramagnetic insulators, whereas the conduction band edge becomes partially occupied in the LN2, LN3, LN5 and LN6 and I3, I4, I5, I8 and I9 structures. The LN4 and I6 structures are ferrimagnetic. The I2 structure had modest electron occupancy in the conduction band edge. The effect of spin polarization – different density-of-state (DOS) function for the two polarization states – was predicted to be stronger in the LN structures. Computations indicated that there is a broad composition range and various cation arrangements which are ferromagnetic. Magnetization maxima were spatially located in the Ti-atoms. This is in line with the computations given in ref. 8. Predictions can be compared with the experiments.

Film $Mg_{1.01}Ti_{1.27}O_3$ has a structure similar to I3, though the occupancy of the 3*a* site is 43%, Table 2. A prediction of magnetization for this structure, weighed by the 3*a* site occupancy factor, is 26emucm$^{-3}$. This is of the same order of magnitude as the experimental values. Film $Mg_{0.73}Ti_{1.38}O_3$ has both Mg and Ti on the basal planes, Table 2, and thus is close to LN-type structures. Due to the small cation fraction on the 3*a* and 3*b* sites the magnetization remains below the values indicated in Fig. 6. When all Vac1 and Vac2 octahedra are filled by Mg and Ti, respectively (see structure LN4 in Fig. 6), the material is predicted to be ferrimagnetic. Mg acts essentially as a donor for a given Ti-O arrangement and the magnetization and resistivity can be adjusted by controlling the occupancy of the conduction band edge states. An important aspect is demonstrated by the structures I2, I3 and LN2: different magnetic properties can be obtained by rearranging the cation distribution within the octahedra, while preserving the stoichiometry. The consequence is that magnetically very different yet structurally nearly identical layers can be manufactured without defected interfaces, which apparently has device applications.

Sample $Mg_{1.49}Ti_{1.03}O_3$ has the ilmenite structure in which the 3*b* sites are filled by Mg, and corresponds to the structure I4 in Fig. 6. The structure is predicted to have two unpaired electrons per unit cell. As discussed above, the experimentally observed zero magnetization is assigned to well-formed Mg-O layers blocking the ferromagnetic coupling. Sample $Mg_{1.24}Ti_{1.17}O_3$ has both Mg and Ti at 3*a* and 3*b* sites and bears a resemblance to the structures I2-I4, three of which are predicted to be ferromagnetic.

The unpaired electrons were dominantly located at Ti, though they generally were not equal, except for the LN6 structure. The fractions of unpaired electrons, when different from zero, are given in Fig. 6. This provides an explanation for the XPS results, Fig. 4(c).

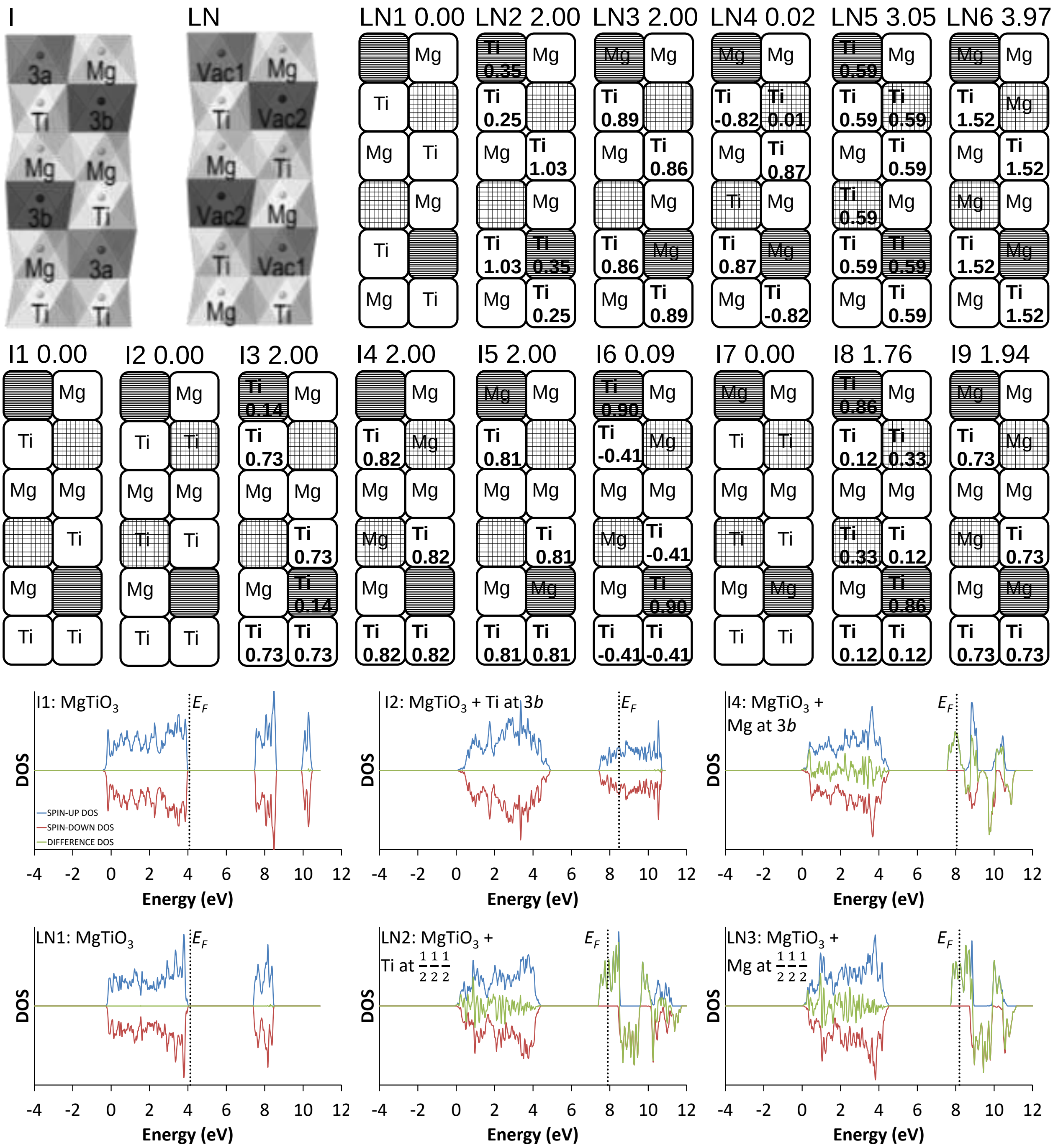

Fig. 6. The structural variants with the corresponding computed magnetization values and selected density-of-state values for up (blue line) and down spins (red line), and their difference (green line). The fraction of unpaired electrons per rhombohedral cell, dominantly localized at Ti, is given.

In summary, ferromagnetism was introduced in both titanium- and magnesium-rich magnesium titanate thin films. By modelling the x-ray diffraction intensities, with constraints set by x-ray photoelectron spectroscopy and Raman scattering measurement, a structural model yielding the level of cation mixing was given. The structure is consisted of an octahedra-network in which the octahedra filling can be controlled. Average titanium valence, magnetization and specific resistivity depend on the cation and vacancy distribution over the octahedral sites. Ferromagnetic samples can be prepared in two variants, with very high resistivity Ti-rich samples and lesser resistive Mg-rich samples. A straightforward model of ferromagnetism, based on the unpaired electron fraction corresponding to the average titanium valence, does not explain the magnetic and electric properties. Band structure computations indicate that the ferromagnetism is based on the partially filled conduction band edge. Partial filling is due to Mg and Ti cations in the octahedra sites vacant in the stoichiometric $MgTiO_3$. The computations show that both paramagnetic and ferromagnetic phases can be obtained for the fixed stoichiometry by varying the cation arrangement in the octahedra sites. The results may find use also in the development of high-quality multiferroics, as ferromagnetism and electric insulation are preferred properties along with the ferroelectricity commonly found in titanium oxides. A second route to introduce ferroelectricity is to intercalate small cations or protons into the double tetrahedral sites present in the ilmenite and corundum structures. If a proton in the double tetrahedron would create a dipole moment, one would have a good chance for ferroelectric material with ferromagnetic properties.

## Acknowledgements

All experimental work was conducted at the Center for Nanophase Materials Sciences, which is a DOE Office of Science User Facility. We thank Dr. Jong Keum (Oak Ridge National Laboratory) for his help with XRD measurements. None of the authors have any competing interests in the manuscript.